\documentclass[preprintnumbers,twocolumn,amsmath,amssymb]{revtex4}

\pdfoutput=1
\usepackage{dcolumn}% Align table columns on decimal point
\usepackage{graphicx}
\usepackage{subfig}
\usepackage{verbatim}
\usepackage{color} 
\usepackage{amsmath}
\usepackage{array}

\begin{document}

\title{Free energy of alternating two-component polymer brushes on cylindrical templates}
\author{William L. Miller}
\author{Behnaz Bozorgui}
\author{Katherine Klymko}
\author{Angelo Cacciuto}
\email{ac2822@columbia.edu}
\affiliation{Department of Chemistry, Columbia University \\3000 Broadway, New York, New York 10027}
\date{\today}
\begin{abstract}
We use computer simulations to investigate the stability of a two-component polymer brush de-mixing
on a curved template into phases of different morphological properties. It has been previously shown via molecular dynamics simulations that immiscible chains having different length and anchored to a cylindrical template will phase separate into stripes of different widths oriented perpendicularly to the cylindrical axis.
We calculate free energy differences for a variety of stripe widths, and extract simple relationships between the sizes of the two polymers, $N_1$ and $N_2$, and the free energy dependence on the stripe width.
We explain these relationships using simple physical arguments based upon previous theoretical work on the free energy of polymer brushes.
\end{abstract}
 
\maketitle
\section*{Introduction}
Polymer brushes are highly tunable systems that have recently attracted 
considerable attention in the scientific community. This is mainly due to the 
already large number of technological applications in which they are used, but also to 
the promising role these materials hold for the future. 
Apart from their well known role in the stabilization of colloidal particles~\cite{Napper}, 
polymer brushes are also used as lubricants, in chromatographic devices and in adhesives,
and their use has recently been proposed in a variety of biotechnological applications including
drug delivery and drug-biocompatibility enhancers~\cite{Dragan,Caster,Ponisseril, Stuart}. Generally speaking, they offer an ideal platform that provides 
control over the physical and chemical properties  of solid and fluid surfaces.
 
Polymer brushes are basically dense systems of polymer chains having one end tethered
to a non-adsorbing surface, and they have been thoroughly studied theoretically 
and numerically on different geometries
(see~\cite{Milner,deGennes,BinderBook,Cotton,Alexander,Halperin,CatesPB, Zhulina,RusselPB,BinderPB} and references therein). The equilibrium properties of these systems 
typically depend on the molecular weight of the chains, their grafting density, and the quality 
of the solvent. Although we have a good qualitative understanding of homogeneous and single components 
brushes, for which scaling arguments have been successfully put forward, the problem becomes 
very complex as soon as we   deal  with non-homogeneous systems with multi-component or nonlinear chains.\cite{vanakaras,linse}

Of particular relevance for the present paper is the work done by Stellacci and 
collaborators~\cite{Stellacci1,Stellacci2,Stellacci3,Stellacci4,Stellacci5} where basically a two-component 
brush of immiscible ligands having different length has been shown to phase separate into striped phases of different widths  when anchored to spherical and cylindrical templates. What is surprising is not the phase separation in itself, but rather the formation of the striped phase and its not well understood dependence on the mismatch in polymer length.
{ Microphase separation is also expected in more traditional  mismatched two-component polymer-brushes~\cite{katz}, where potential applications in stimuli-responsive systems and nanotemplating have recently attracted great attention 
to the field (see ref~\cite{ayres} and references therein).}
Being able to control pattern formation at the nanoscale is a core element in the production 
of novel materials via the process of self-assembly. 

Direct coarse grained molecular dynamics simulations by Glotzer et al. have clearly indicated  
the driving forces behind the formation of these phases as the difference between the chains molecular weight, and the overall 
height of the brush itself~\cite{Stellacci2,Stellacci4,Stellacci5,Glz1,Stellacci6}.
{ Very recently, a coarse-grained model for two-component polymer mixtures of different length has also been put forward, and 
the formation of different phases as a function of length mismatch on a planar geometry has been analyzed, revealing the 
presence of striped phases in these systems as well~\cite{katz}}.
 
Unlike other numerical work that focused on the details of the de-mixing transition,   
in this paper { we focus exclusively on the origin of the microphase separation}.
We use numerical simulations to compute free energies of an  
inhomogeneous two-component polymer brush to understand how molecular weight and overall brush  height
affect the stability and width of the striped phases. 
Specifically, we focus on the case of a cylindrical solid template where stripes have been shown to form
promptly both numerically and experimentally~\cite{Stellacci6,Stellacci5}, and we compute, at constant grafting density  
and template radius, how the system free energy changes when imposing stripes of different width and height on the template.
{ Our results indicate that what limits macrophase separation into a two-phase region, thus leading to the stabilization of the striped phase, is 
the elastic strain that  builds up in the polymer brush formed by the mismatched (exposed) polymer segments.}
   
\section*{Methods}

To ensure that our data are not  affected by the lateral tension due to the 
immiscibility of  the two polymer types, and that we are indeed measuring exclusively 
the free energy difference arising from the chains' length mismatch and overall brush morphology, 
we simulate systems in which the only difference between the two polymer types is their length. 
This is equivalent to setting the  line tension between 
the stripe's boundaries that comes directly from the immiscibility term in the pair 
potential between the two polymer types equal to zero.
As a consequence, we are required to lock in place the position of the anchoring monomers to prevent the different chains from trivially mixing. 
 
\begin{figure}
	\subfloat{
		\includegraphics[width=0.2\textwidth]{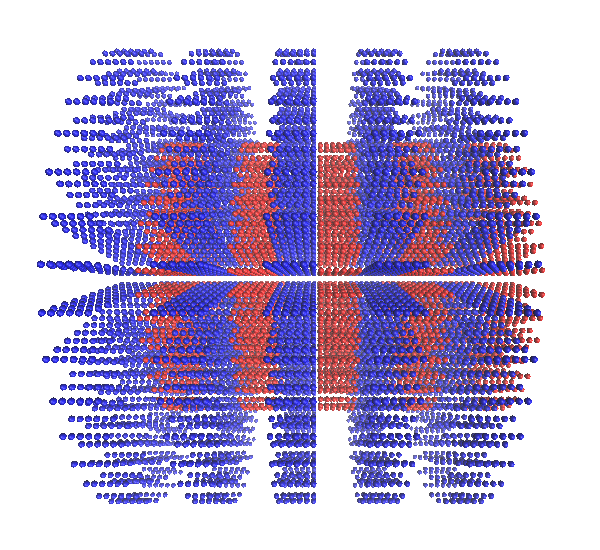}
	}
	\subfloat{
		\includegraphics[width=0.2\textwidth]{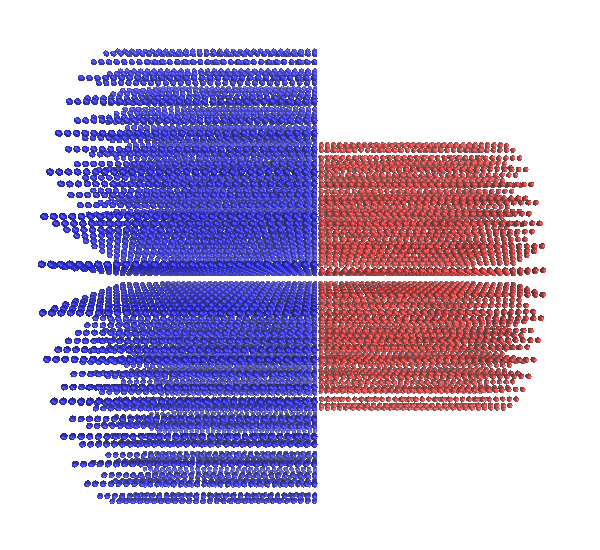}
	}
	\caption{Snapshots of typical initial configurations for $L_p = 6$ and $L_p = 30$, respectively.}\label{snap}
\end{figure}

Polymers are modeled as  sequences of spherical beads of radius $\sigma$ linearly connected via a harmonic potential $V_{\textrm{bond}}(r) = \kappa \left(\frac{r}{\sigma}-1\right)^2$, with $\kappa = 800 k_{\rm B}T$.
Any two monomers in the system interact 
via the soft and purely repulsive dissapative particle dynamics (DPD) simulation potential:
\[V^{ij}(r_{mn}) = 
	\begin{cases}
		\epsilon_{ij} \left(1-\frac{r_{mn}}{\sigma}\right)^2 & \textrm{if $r \leq \sigma$} \\
		0 & \textrm{otherwise}
	\end{cases}
\]
where $i$ and $j$ indicate the identity of the polymer $i,j \in\{1,2\}$ and $m,n \in\{1,N_i\}$ refers to the identity of the monomer. 
Here $N_i$ is the length of a polymer of type $i$. 
All the results that follow use $\epsilon_{11} = \epsilon_{22} = \epsilon_{12} = 5 k_BT$.
Each polymer is grafted to a fixed point on the outer surface of a cylindrical template of  radius $R=2.5\sigma$ via the same harmonic potential tethering the consecutive monomers in a chain. In all our simulations we considered a total number of polymers $n_p=2580$ arranged in a homogenous grid having square symmetry. The polymer identity is finally selected to generate alternating stripes of  width $L_p$.
For a lateral grafting density of the polymers equal to $2.74/\sigma$, our system contains $60$ one-polymer-wide rings, making the possible values of $L_p\in \{1, 2, 3, 5, 6, 10, 15, 30\}$, when equal number of polymer types are considered $n_p^{(1)}=n_p^{(2)}=n_p/2$. Figure~\ref{snap} shows snapshots of typical initial configurations 
for $L_p=6$ and $L_p=30$. { This particular geometry is selected because we find that test unconstrained 
simulations of immiscible chains always lead to stripe formation perpendicular to the cylindrical axis.} This result has also been observed  
in~\cite{Stellacci6}.
 
Given the size of the system we carried out our simulations using the  molecular dynamics package  {\sc LAMMPS}~\cite{lammps} with Langevin dynamics in the $NVT$ ensemble. Dimensionless units are used throughout this paper. The timestep size was set to $dt=0.0025\tau_0$ ($\tau_0$ is the dimensionless time).

Our goal is to compute the free energy difference $\Delta F(L_p)$ between systems having different stripe width $L_p$, while keeping
everything else unaltered, as a function of the difference between the lengths $N_1$ and $N_2$. To compute the system free energies 
we use  the thermodynamic integration method~\cite{FrenkelBook}.
The idea is to introduce a fictitious potential 
\[V^{ij}_{\lambda}(r_{mn}) = \lambda
	\begin{cases}
		\epsilon_{ij} \left(1-\frac{r_{mn}}{\sigma}\right)^2 & \textrm{if $r \leq \sigma$} \\
		0 & \textrm{otherwise}
	\end{cases}
\]
which acts between any two monomers in the system. For $\lambda=0$ the chains are ideal and the system free energy is independent of
the specific grouping of the polymer types;  as $\lambda\rightarrow 1$ we recover the system of interest. 
The free energy of the full system can then be extracted by performing the following integral 
\begin{equation}
 F(L_p)=\int_0^{1}d\lambda \left(\frac{d V^{ij}_{\lambda}}{d\lambda}\right)_{\lambda}
 \end{equation}
In practice, we perform simulations for several values of $\lambda$, and numerically compute the integral above. 

\section*{Results}

\begin{figure}
	\subfloat{
		\includegraphics[width=0.4\textwidth]{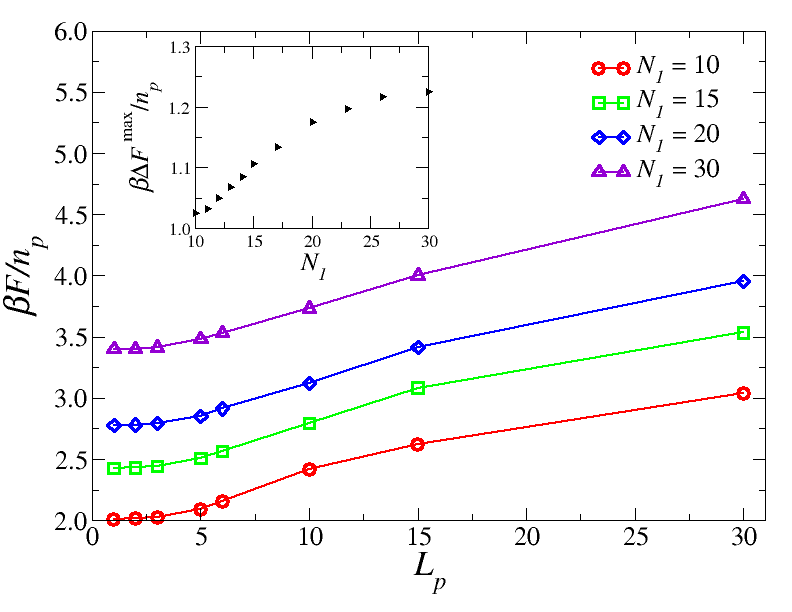}
		\put(-205,128){\subref{FvsLp}}
		\label{FvsLp}
	} \\
	\subfloat{
		\includegraphics[width=0.4\textwidth]{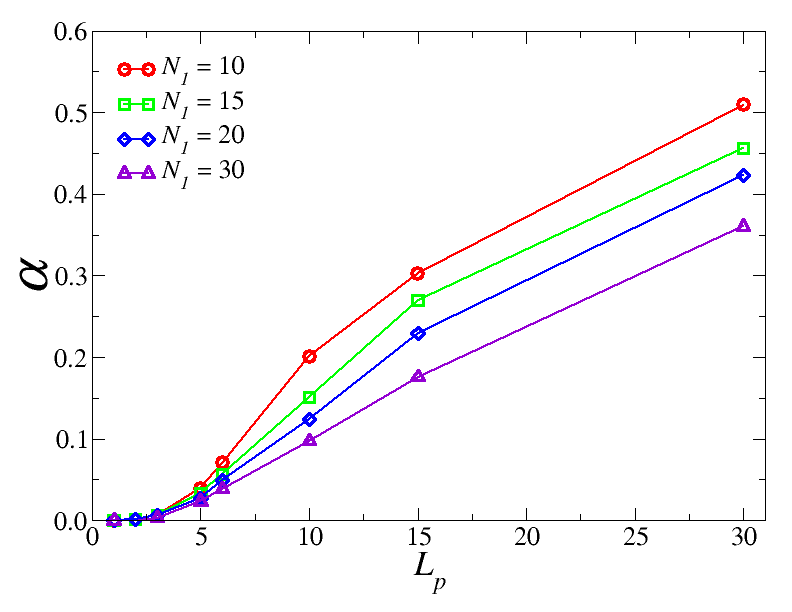}
		\put(-205,128){\subref{FvsN}}
		\label{FvsN}
	} \\
	\caption{\protect\subref{FvsLp} Free energy per polymer $F/n_p$ vs. $L_p$ for a cylinder of radius $2.5\sigma$, for various values of $N_1 \in \left[10,30\right]$ with $N_2 = 0$.  Each line represents a different value of $N_1$; increasing $N_1$ leads to an increase in free energy $F$ at every value of $L_p$. The inset shows how the free energy gap $\Delta F^{\rm max}$ defined as the free energy difference between $L_p=30$ and  $L_p=1$
depends on the overall length of the chains $N_1$. \protect\subref{FvsN} The same data plotted as $\alpha = \frac{F(L_p)-F(1)}{F(1)}$ vs. $L_p$.}
\end{figure}

To get insight into the problem we first consider the case in which $N_2=0$; that is, only one chain type is present in the system. Figure~\ref{FvsLp} shows our numerical data for the free energy cost  $F(L_p)$ associated with the partitioning of the two polymer species into alternate stripes of different $L_p$ on a cylinder of radius $2.5\sigma$ and lateral grafting density $2.74/\sigma$, as a function of $L_p$ and for polymer lengths $N_1 \in \left[10,30\right]$ and $N_2 = 0$.
  This specific lateral grafting density was selected because unconstrained simulations under this condition lead to prompt microphase separation for moderate values of polymer immiscibility. Furthermore, a sufficiently large value of lateral grafting density guarantees large enough differences in free energies as a function length mismatch.
  
Our data show that the most favorable state corresponds to that having the largest number of stripes (i.e. the smallest value of $L_p$). The inset on the same figure shows how the free energy difference $\Delta F^{\rm max}\equiv F(L_p=30)-F(L_p=1)$ grows with the length of the chain $N_1$.
 These results can be qualitatively understood by referring back to the theory of polymer brushes on flat surfaces and its extension to spherical and cylindrical  
surfaces~\cite{Cotton,LipowskyPB}. The key point is that the main contribution to the free energy cost per chain associated with a polymer brush has an inverse dependence on the the lateral chain-to-chain grafting distance $D$. 
\begin{equation}
F_{\rm cyl}\sim 2 k_{\rm B}T \left (\frac{R}{D}\right )\left [\left (1+\frac{4h_0}{3R}\right )^{3/8}-1\right ]
\label{F_cyl}
\end{equation} 
where $h_0\sim \sigma N D^{-2/3}$ is the height of a brush made with polymers of length $N$ grafted on a planar surface. Although in our system the location of the grafting lattice on the cylindrical surface 
is independent of $L_p$ and  therefore the total grafting density of the system is kept constant at all times, 
different distributions of the chains on the lattice result in different local densities. 
Crucially, when considering the free energy contribution due to the different lateral organization of the chains, one can consider our system with $L_p=1$ as a polymer brush with twice the lateral grafting distance of that relative to $L_p=30$. Clearly the latter system cannot be straightforwardly described with Eq.~\ref{F_cyl} due to the free boundaries of the brush, nevertheless we expect that for $L_p\rightarrow\infty$ 
the role of the boundary should become negligible. If we now introduce the dimensionless parameter
$\alpha (L_p)$ defined as
\begin{equation}
\alpha\equiv\frac{F(L_p)-F(1)}{F(1)},
\end{equation}
the argument above leads to clear  upper-bounds on the value of $\alpha$.
Specifically, we expect $\alpha_{\rm max}=2^{5/3}-1$ in the flat limit, and $\alpha_{\rm max}=2^{5/4}-1$ for $(h_0/R)\gg1$,  independently of chain length. Our numerical results are indeed within these bounds.
 
\begin{figure}
	\subfloat{
		\includegraphics[width=0.2\textwidth]{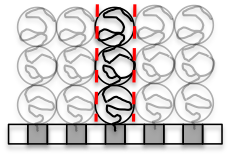}
	}
	\subfloat{
		\includegraphics[width=0.22\textwidth]{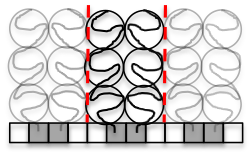}
	} \\
	\subfloat{
		\includegraphics[width=0.3\textwidth]{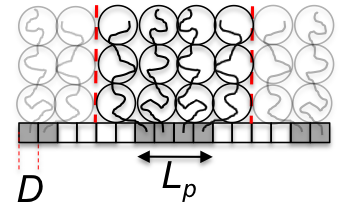}
	}
	\caption{Sketch of expected chain configurations for $L_p = 1, 2,$ and $4$.}\label{geom}
\end{figure}

Figure~\ref{FvsN} shows $\alpha$ plotted as a function of $L_p$ for different values of $N_1$.
$\alpha$ has a nontrivial dependence on $L_p$; nevertheless, 
simple geometrical arguments can be used to understand at least qualitatively the overall
behavior of these curves. Figure~\ref{geom} shows a sketch of the expected chain distributions 
for $L_p=1,2$ and $4$ for the simpler case of a flat surface.
For small values of $L_p$ the problem is dominated by boundary effects, and we can imagine 
the chains equally sharing the overall space $2L_p$ per stripe available to them.

It is immediately obvious that the free energy difference between $L_p=1$ and $L_p=2$ per chain should 
be very small indeed. In both cases the chains can be accommodated into space as a sequence 
of blobs of diameter equal to twice the grafting distance $D$.  A similar argument can be made for $L_p=4$. 
The main extra cost in free energy per polymer for a growing $L_p$ should come from the weak entropic stretching of the polymers next to the grafting surface and should weakly grow as $L_p^2$.

When $L_p$ becomes large, 
the free energy of the system is dominated by chains 
in the bulk of the  stripe. In this regime the blob size of the outer chains is larger than that of the inner chains.
Assuming a generic size profile for the blobs of the form $D(i)=D(1+ c \,i^{\beta})$ where $i\in [1,L_p]$ and
 $c$ is fixed by the constraint $\sum_1^{L_p} D(i)=2D_0L_p$, 
one can estimate the free energy per stripe as $F_s\simeq k_{\rm B}T N_1 \sigma^{1/\nu} \sum_1^{L_p}  \frac{1}{D(i)^{1/\nu}}$. Given that there are $n_p/L_p$ particles in each stripe, the free energy per particle on a flat interface  reads as
\begin{equation}
F\simeq k_{\rm B}T N_1 \frac{\sigma^{1/\nu}}{L_p} \sum_1^{L_p}  \frac{1}{D(i)^{1/\nu}}
\end{equation}
It is difficult to extract $\beta$ from our simulations, but from visual inspection of the brush profiles 
one should  expect $\beta\geq2$, leading  to functional forms for $F$ having the typical saturating behavior 
shown in our data. 

In the current form the data do not  collapse onto a universal master curve. In fact, the free energy difference of shorter chains tends to grow faster than that of the longer ones. 
This result is quite revealing as it directly reflects on the 
nontrivial interactions between neighboring regions of grafted polymers.

\begin{figure}
	\includegraphics[width=0.4\textwidth]{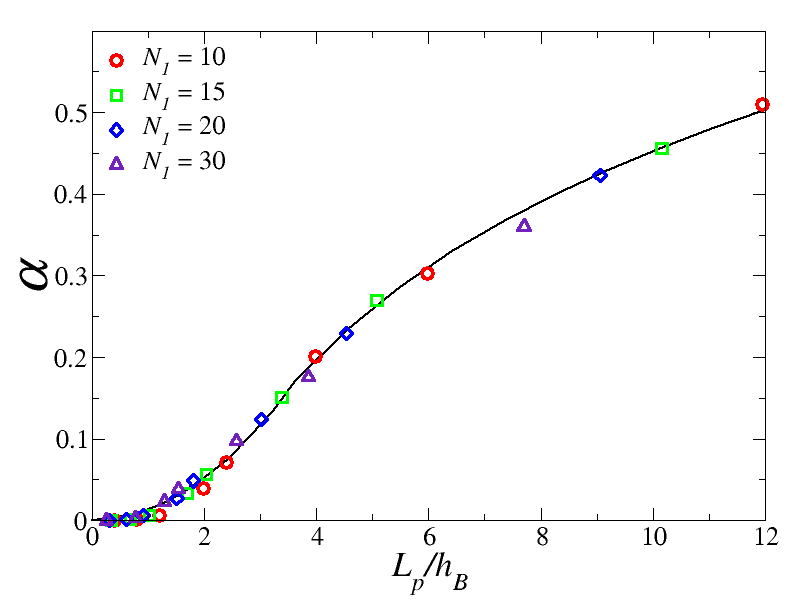}
	\caption{$\alpha$ vs. the rescaled $L_p/h_B$.  The black line is a guide to the eye.}\label{FvsLpscaled}
\end{figure}

Such interactions will cease as soon as the mutual distance between two neighboring regions is larger than
twice the side spread of the polymers at the boundaries. Clearly this will happen sooner for the 
short polymer than for the longer ones. We should be able to account for this difference by  
rescaling of $L_p$ with  $L_p/N_1^{\gamma}$.
We find the best collapse for $\gamma=0.4$. This result suggests a 
length-scale, $h_B$, controlling the extent of the side interactions  that scales as $h_B=\sigma h_0/R_G$. 
Figure~\ref{FvsLpscaled} shows the data collapse into a universal master curve $\alpha={\tilde \Phi}(\frac{L_p}{h_B})$.

Let's now turn our attention to the case in which $N_2\neq 0$.
It would be desirable to develop a simple way of extending our results to the more 
general $N_2 \neq 0$ case. In particular, we are interested in understanding whether it is possible
to map a system having different polymer  lengths $N_1 > N_2 \neq 0$ to a system of a single polymer type of length $N_1^\prime = N_1-N_2$ (and $N_2^\prime = 0$.)

\begin{figure}
	\includegraphics[width=0.4\textwidth]{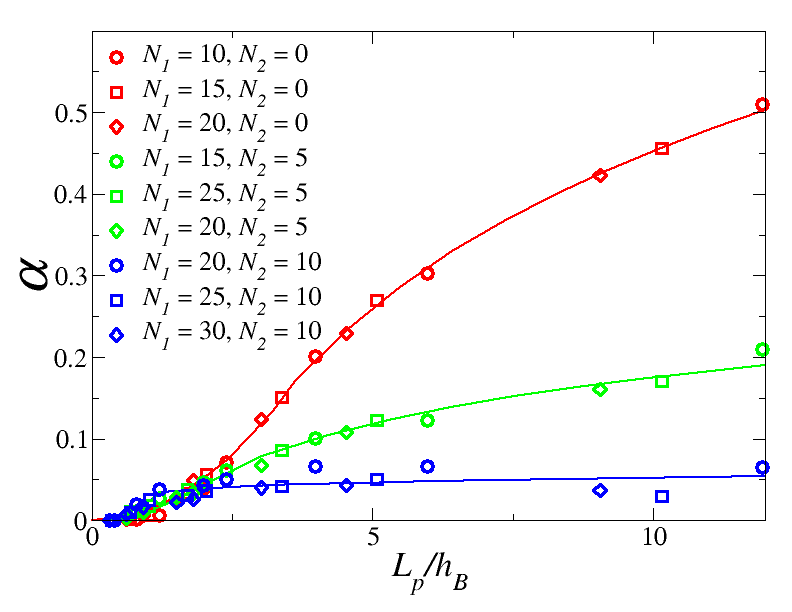}
	\caption{$\alpha$ vs. the rescaled $L_p/h_B$ for different values of $N_2$.  The lines are guides to the eye.}\label{FvsLpscaleddiff}
\end{figure}

Figure~\ref{FvsLpscaleddiff} shows $\alpha$ as a function of $L_p/h_B$ for different values of $N_1$ at 
$N_2=5$ and $N_2=10$. The results for $N_2=0$ are also shown as a reference.
As the cause for the free energy gap between the different configurations 
rests in the difference  $N_1-N_2$, to properly compare the data coming from systems having 
different values of $N_2$, when computing $\alpha$
we subtracted from the measured free energies $F(L_p)$ the constant 
core free energy of the underlying full brush of height $N_2$, $F_{\rm core}(N_2)$.
When $N_2\neq0$, we define $\alpha$ as  $\alpha(L_p)=(F(L_p)-F(1))/(F(1)-F_{\rm core}(N_2))$.

It is comforting  to report that indeed it is possible to systematically collapse all of the data corresponding to 
different values of $N_1$ at a given (fixed) $N_2$ into unique curves;
however, we cannot collapse data coming from different values of $N_2$. This result clearly points to the fact that the universal curve previously described for $N_2=0$ also has a nontrivial dependence on $N_2$: $\tilde \Phi=\tilde \Phi (\frac{L_p}{h_B},N_2)$. 
In principle, this shouldn't come as a  surprise for brushes grafted on a curved surface,
in fact, the effect of the ``interior'' layer of $N_2$ monomers is to increase the effective radius of the cylinder to which the reduced system would be grafted onto, and to decrease the grafting density of the effective brush of length $N_1-N_2$; both are variables the free energy depends on (see Eq.~\ref{F_cyl}).

Simple geometrical arguments can be used to estimate the change in lateral grafting distance.
Following~\cite{LipowskyPB}, we can write the change in lateral grafting  spacing
as we radially move away from the cylinder surface of an amount $r$ as $\xi(r)=D (r/R)^{1/2}$. 
For short polymers and 
at the relatively high densities of our systems we have $r\simeq R+\sigma N_2$. It follows that
$\xi(r)=D (1+\sigma N_2/R)^{1/2}$. Simultaneously, the effective polymer of length $N_1-N_2$ 
would be grafted onto a cylinder of radius $R^\prime\simeq R+\sigma N_2$. The ratios $R^\prime/\xi$
appearing in the brush free energy should then be adjusted to
 $R^\prime/ \xi(r) \rightarrow (R/D) (1+\sigma N_2/R)^{1/2}$  introducing a predictable dependence on $N_2$
 in the problem. Although this argument suggests an overall increase in free energy for the system at a given $L_p$, 
 it isn't obvious how it will affect the dimensionless ratio $\alpha (L_p)$. Clearly this correction should factor out in the large $R$ limit or for $h_0/R\gg1$.
 
Our numerical data show that $\alpha$ is quite sensitive to $N_2$. Specifically,  
$\alpha(L_p)$ tends to flatten as $N_2$ increases. 
This is most likely explained by the fact that the base of the effective chain of length $N_1-N_2$
is not really grafted in place, but the layer of polymers below allow for significant reorganization of the base
due, for instance, to their compressibility. Because of the blob-size dependence
on the distance from the surface, on a convex template the chains become systematically more laterally compressible
as $N_2$ is increased. 
The net result is a flattening of the free energy difference as a function of $L_p$.
Unfortunately, we find that our data for different values of $N_2$
cannot be collapsed by a simply rescaling of the free energy with a power of $N_2$,
indicating a subtle interplay between $F(L_p)$ and the height of the bottom layer.

\section*{Conclusions}

In summary, we analyzed the relative stability of the striped phases  arising from the de-mixing of immiscible polymers 
on a cylindrical template as a function of the width of the stripe, the length mismatch between the chains, and the overall length of the brush. We set numerical bounds for the free energy gap between an alternating (many thin stripes) and a fully de-mixed (two wide stripes) phase, and found that indeed the former
becomes more and more favorable as the mismatch between the chains' lengths, $\Delta N$ is increased.
We also found that when appropriately normalized the free energies as a function of stripe width for different values of $\Delta N$  can be collapsed into unique master curves that only depend on the overall length of the lower brush $N_2$. Finally we discussed the possibility of reducing a two-component system of chains 
having lengths $N_1>N_2$ into a simpler single component system of chains of length $\Delta N$.
Our results are fully consistent with previous molecular dynamics simulations on this problem
{ in identifying the key parameters setting the width of the stripes in the polymer length mismatch $N_1-N_2$, 
and the overall length of the shorter brush $N_2$ together with the degree of chain immiscibility}. 

In all our  data the direct line tension $\gamma$ coming from the pair potential between chains of different types 
has been set to zero. This term would add to the free energy balance a predictable contribution of the type
$F_l(L_p,N_2)\simeq\gamma \, h_0(N_2) (N_p/L_p)$, where $h_0(N_2)$ is the height of the polymer brush formed by the lower chains and $(N_p/L_p)$ is the number of phase boundaries. Clearly $F_l(L_p,N_2)$ has a minimum for 
$L_p\rightarrow\infty$ and is large for small $L_p$. The balance between the line tension and the configurational free energy computed in this paper should therefore set the width of the stripe. More work in this direction is currently
underway.

\section*{ACKNOWLEDGMENTS}
This work was supported by the American Chemical Society under PRF grant No. 50221-DNI6.

\end{document}